\documentclass{PoS}
\title{Comparing Tensor Renormalization Group and Monte Carlo calculations for spin and gauge models}

\ShortTitle{Comparing Tensor Renormalization Group and Monte Carlo}

       \author{\speaker{Yannick Meurice}$^a$\footnote{Current email address: yannick-meurice@uiowa.edu},
       Alan Denbleyker $^a$, Yuzhi Liu $^{a,c,d}$, Tao Xiang$^b$,  Zhiyuan Xie$^b$,  Ji-Feng Yu$^b$, Judah Unmuth-Yockey$^a$, Haiyuan Zou $^a$\\
\llap{$^a$} Department of Physics and Astronomy, University of Iowa, Iowa City, IA 52240, USA\\
\llap{$^b$} Institute of Physics, Chinese Academy of Sciences, P.O. Box 603, Beijing 100190, China\\
\llap{$^c$} Fermi National Accelerator Laboratory, Batavia, IL 60510, USA \\
\llap{$^d$} Department of Physics, University of Colorado, Boulder, CO 80309, USA\\
}

\abstract{We show that the Tensor Renormalization Group (TRG) method can be applied to O(N) spin models, principal chiral models and pure gauge theories (Z2, U(1) and SU(2)) on (hyper) cubic lattices. 
We explain that contrarily to some common belief, it is very difficult to write compact formulas expressing the blockspinning of lattice models. 
We show that in contrast to other approaches, the TRG formulation allows us to write exact blocking formulas with numerically controllable truncations.  
The basic reason is that the TRG blocking separates neatly the degrees of freedom inside the block and which are integrated over, from those kept to communicate with the neighboring blocks.
We argue that the TRG is a method that can handle large volumes, which is crucial to approach quasi-conformal systems. The method can also get rid of some sign problems.
We discuss recent results regarding the critical properties of the 2D O(2) nonlinear sigma model with complex $\beta$ and chemical potential.
As some of these results appeared in a recently published paper (PRD 88, 056005) and two recent preprints  (arXiv:1309.4963 and  arXiv:1309.6623), these proceedings rather emphasize the 
conceptual aspects of our ongoing effort.}

\FullConference{31st International Symposium on Lattice Field Theory LATTICE 2013\\
		 July 29-August 3, 2013\\
		 Mainz, Germany}

\begin{document}

\section{Introduction: blocking in configuration space is hard! }
The block spin idea played a central role in the development of the RG method by the late Ken Wilson. However, successful applications of the method were possible without requiring a numerical implementation 
of the original idea. Blocking in configuration space is hard! Anyone who believes that blocking is a straightforward procedure should attempt to write a simple algorithm for the two-dimensional Ising model on a square lattice.
For this model, blocking means replacing four spins in a 2x2  square block by a single variable and write an effective energy function (or at least some effective measure) for the new block variables. 

A possible strategy is to consider 2 by 2  blocks in  a $A-B$ checkerboard. First, we consider the $B$ blocks as fixed backgrounds.  We can then block spin in the $A$ blocks. For instance, 
 $Prob(\phi_A=4| \phi^{background}_i)\propto \exp(\beta(4+\sum_{i=1}^8 \phi_i^{background}))$ and more complicated expressions for the intermediate values of $\phi_A$. So far, we have blocked 1/2 of the total number of spins.
 The next step is to try to block spin in the $B$ blocks. Using the results of step 1 we can block spin in a given $B$ block surrounded by 4 $A$ blocks but this set of 5 blocks is surrounded by 20 background spins. We have blocked 
 5/8 of the total number of spins but the procedure becomes more intricate at each step and finding the effective energy function is nontrivial. This is explained in more detail and illustrated in \cite{Exactblocking13prd}. The goal of this discussion is only to illustrate the 
 generic difficulty of blocking in configuration space. 

It is possible to invent approximations where no new interactions are generated by the blocking process. Well-known examples are the Migdal-Kadanoff approximation \cite{Migdal:1975zf,Kadanoff:1976jb}, the approximate recursion formula \cite{wilson71b}  or other hierarchical approximations \cite{baker72,hmreview}. However, in these examples, the lack of reference to an exact procedure makes the systematic improvement of these approximations difficult. 
In contrast, the TRG formulation allows us to write exact blocking formulas with numerically controllable truncations.  
The basic reason is that the TRG blocking separates neatly the degrees of freedom inside the block (which are integrated over), from those kept to communicate with the neighboring blocks.
The TRG approach of classical lattice models was introduced in Refs. \cite{JPSJ.64.3598,PhysRevLett.99.120601,PhysRevB.79.085118} motivated by 
 tensor states developed in RG studies of quantum models \cite{uli}.  Improved methods to take into account the environment were proposed in Refs.  \cite{PhysRevLett.103.160601,PhysRevB.81.174411,PhysRevB.86.045139}. In Ref. \cite{Exactblocking13prd}, we showed that 
TRG methods can be applied to models studied by lattice gauge theorists, namely spin models ($O(N)$ and principal chiral models) 
and pure gauge models (Ising, $U(1)$ and $SU(2)$). The case of actions quadratic in Grassmann variables is briefly discussed in \cite{YMPRB13}. 

In these proceedings, we review the general procedure with the simple example of the 2D Ising model (Sec. \ref{sec:simple}). More complicated examples (2D $O(2)$ model and 
3D Abelian gauge theories) are discussed in Sec. \ref{sec:examples}. The TRG method has the potential to handle situations where there is a sign problem. We briefly discuss recent 
results for spin models at complex $\beta$ and with chemical potential in Sec. \ref{sec:trgsign}. More details can be found in recent preprints\cite{trgsign,trgo2}. 

\section{TRG blocking: it's simple and exact!}
\label{sec:simple}
\begin{figure}[h]
       \begin{center}
            \includegraphics[width=2.in,angle=0]{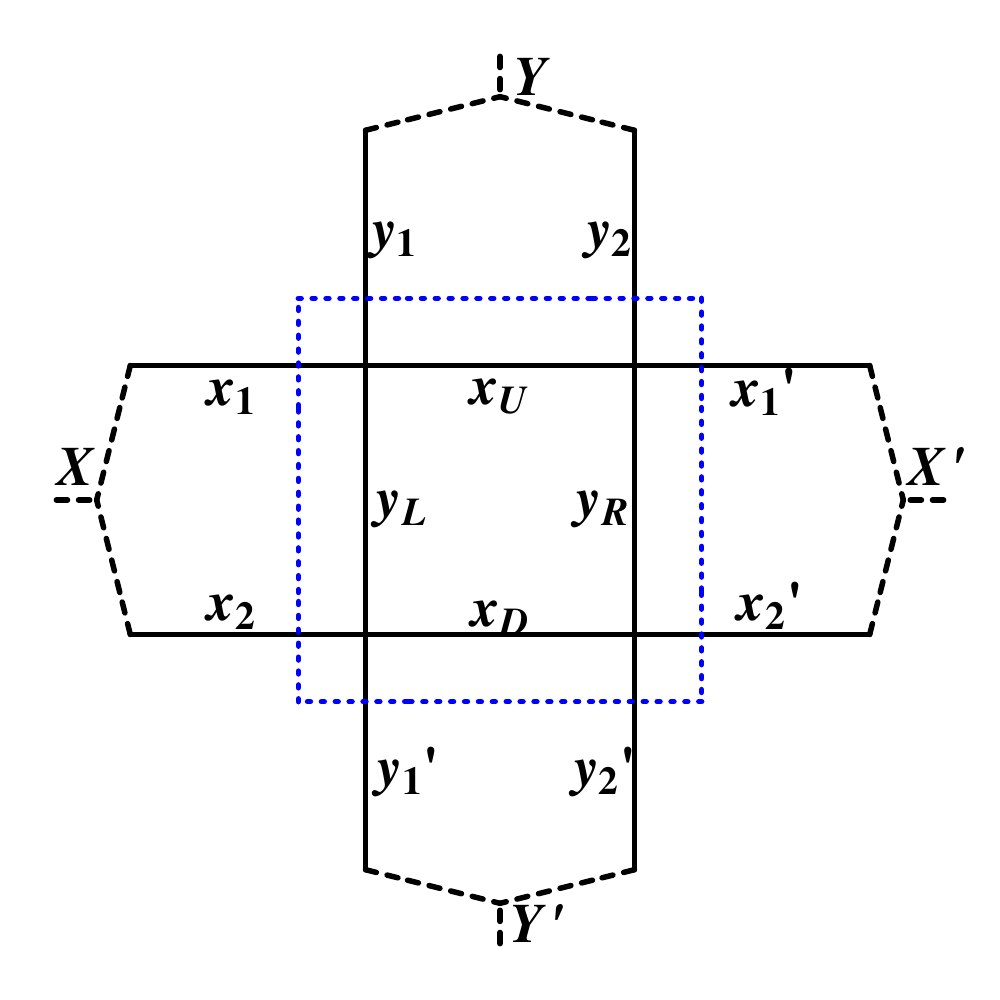}
            \caption{Graphical representation of $T'_{XX'YY'}$ . 
            \label{fig:block} }
           \end{center}
        \end{figure}

The first step in the TRG formulation of spin models is to use character expansions and assign degrees of freedom to the links. 
This first step is similar to what is done when we construct dual formulations, however, we do not introduce dual variables here. 
In the case of the Ising model, we write for each link:
        \begin{eqnarray}
       && \exp(\beta \sigma_1 \sigma _2)=\cosh(\beta)(1+\sqrt{\tanh(\beta)}\sigma_1    \sqrt{\tanh(\beta)}\sigma_2 )  =  \nonumber \\
        &&cosh(\beta) \sum _{n_{12}=0,1} (\sqrt{\tanh(\beta)}\sigma_1    \sqrt{\tanh(\beta)}\sigma_2)^{n_{12}} .
        \label{eq:char}
        \end{eqnarray}
 We then regroup the four terms involving a given spin $\sigma_i$ and sum over its two values $\pm 1$. The results can be expressed in terms of a tensor:
 $T^{(i)}_{xx'yy'}$ which can be visualized as a cross attached to the site $i$ with the four legs covering half of the four links attached to $i$. The entire lattice 
 can be tiled with such crosses. 
  The horizontal indices $x,\ x'$ and vertical indices $y,\ y'$ take the values 0 and 1 as the index $n_{12}$. 
\begin{equation}
            T^{(i)}_{xx'yy'} =f_x f_{x'}f_y f_{y'} \delta\left(\rm{mod}[x+x'+y+y',2]\right) \ ,
            \label{eq:factor}
        \end{equation} where $f_0=1$ and $f_1 =\sqrt{ \tanh(\beta)}$. The delta symbol is 1 if $x+x'+y+y'$ is zero modulo 2 and zero otherwise. 
 The partition function can be written exactly as
 \begin{equation}            Z = (\cosh (\beta))^{2V}Tr \prod_{i}T^{(i)}_{xx'yy'}\  ,
 \end{equation}
 where $Tr$ means contractions (sums over 0 and 1) over the links. 
 This representation reproduces the closed paths of the high temperature  expansion.  
 TRG blocking separates the degrees of freedom inside the block which are integrated over, from those kept to communicate with the neighboring blocks. 
 This is represented graphically in Fig. \ref{fig:block}.
        
Blocking defines a new rank-4 tensor $T'_{XX'YY'}$ where each index now takes four values. Its explicit form is:
        \begin{eqnarray}
            \label{eq:square}
            &\ &T'_{X({x_1},{x_2})X'(x_1',x_2')Y(y_1,y_2)Y'(y_1',y_2')} = \\ \nonumber
            &\ &\sum_{x_U,x_D,x_R,x_L}T_{x_1 x_U y_1y_L}T_{x_Ux_1'y_2y_R}T_{x_Dx_2'y_R y_2'}T_{x_2x_Dy_Ly_1'}\  ,
        \end{eqnarray}
        where $X(x_2,x_2)$ is a notation for the product states e. g. ,
        $X(0,0)=1,\  X(1,1)=2, \  X(1,0)=3,\  X(0,1)=4$. 
 The partition function can now be written exactly as 
        \begin{equation}
            Z=(\cosh (\beta))^{2V}Tr\prod_{2i}T'^{(2i)}_{XX'YY'} \ , 
            \label{eq:ZP}
        \end{equation}
where $2i$ denotes the sites of the coarser lattice with twice the lattice spacing of the original lattice. 
The new expression has exactly the same form as the previous one except for the fact that the indices run over more values. 

In practice, this exact procedure will require too much memory after a few iterations and one needs to introduce truncations. 
For the Ising model on square and cubic lattices,  a truncation method  based on Higher Order SVD (HOTRG) gives very accurate results \cite{PhysRevB.86.045139} . 
It was also noticed that  the Higher Order SVD \cite{YMPRB13}
sharply singles out a surprisingly small subspace of dimension two. 
In the two states limit, the transformation can be handled analytically yielding a value 0.964  for the critical exponent $\nu$ much closer to the exact value 1 than 1.338 obtained in Migdal-Kadanoff approximations.  Alternative blocking procedures that preserve the isotropy can improve the accuracy to $\nu=0.987$ and 0.993.  

\section{More examples}
\label{sec:examples}
The previous results can be generalized for the $O(2)$ model. The partition function reads
                       \begin{equation}
            Z = \int{\prod_i{\frac{d\theta_i}{2\pi}} {\rm e}^{\beta \sum\limits_{<ij>} \cos(\theta_i - \theta_j)}}.
            \label{eq:bessel}
        \end{equation}
For each link, we can use
        \begin{equation}
            {\rm e}^{\beta  \cos(\theta_i - \theta_j)} = \sum\limits_{n_{ij}=-\infty}^{+\infty} {\rm e}^{i n_{ij}(\theta_i-\theta_j)} I_{n_{ij}}(\beta)\  ,
        \end{equation}
        where the $I_n$ are the modified Bessel functions of the first kind. 
         In two dimensions, we obtain the factorizable expression for the initial tensor:
        \begin{equation}
            T^i_{n_{ix},n_{ix'},n_{iy},n_{iy'}} = \sqrt{I_{n_{ix}}(\beta)} \sqrt{I_{n_{iy}}(\beta)} \sqrt{I_{n_{ix'}}(\beta)} \sqrt{I_{n_{iy'}}(\beta)}
           \delta_{n_{ix}+n_{iy},n_{ix'}+n_{iy'}} \ .
        \end{equation}
        The partition function and the blocking of the tensor are similar to the Ising model. The only difference is that the sums run over the integers.  
        As the $I_n (\beta)$ decay rapidly for large $n$ and fixed $\beta$ (namely like $1/n!$) there is no convergence issue. 
        The generalization to higher dimensions is straightforward. 
        In Ref. \cite{Exactblocking13prd}, exact blocking formulas are also provided for the $O(3)$ model and the $SU(2)$ principal chiral model. 
       
       The TRG method can also be used to handle gauge theories. We briefly describe the so-called symmetric formulation for Abelian groups of Ref. \cite{Exactblocking13prd} where the $SU(2)$ case is also discussed.
            A blocking procedure can be constructed by sequentially combining two cubes into one in each of the directions. 
                        \begin{figure}[h]
                \begin{center}
                    \includegraphics[width=7cm]{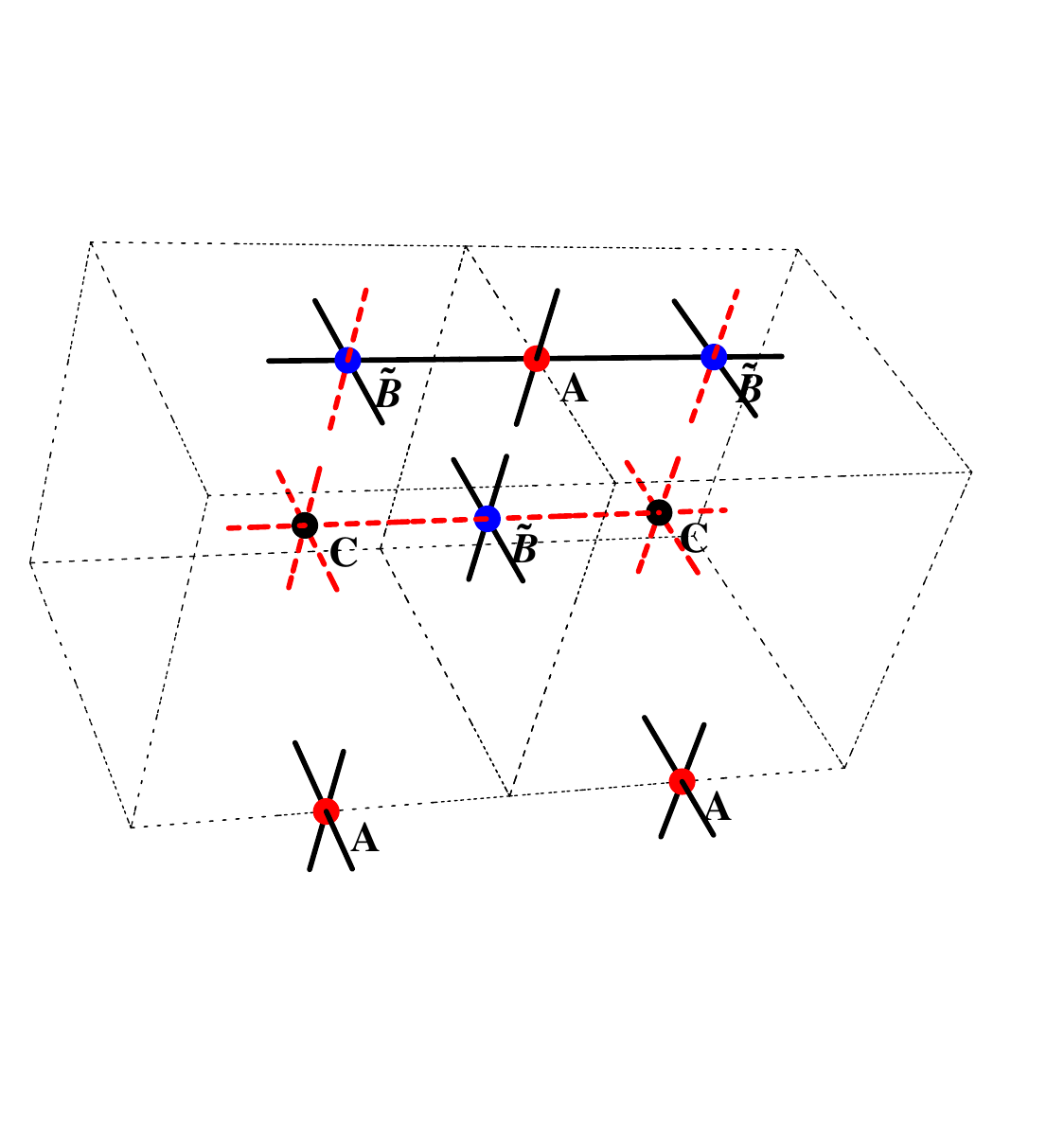}
                    \vspace{-1.1cm}
                \end{center}
                \caption{\label{fig:isotc} Illustration of the blocking procedure}
            \end{figure}
            On the links in the blocked direction of the new lattice formed by two cubes, two parallel $A$ tensors form the new $A'$ tensor with product states (capital letters). Each tensor element can be written as
            \begin{eqnarray}
                \label{eq:newa1}
               {A'}_{X(x_1,x_2)X'(x_1',x_2')Y(y_1,y_2)Y'(y_1',y_2')}=A_{x_1x_1'y_1y_1'}\times A_{x_2x_2'y_2y_2'}.
            \end{eqnarray}
            On the new faces, two $\tilde{B}$ tensors and one $A$ tensor form a new $\tilde{B}'$ tensor,
            \begin{eqnarray}
                \label{eq:newb1}
                \nonumber
               {\tilde{B}'}_{xx'Y(y_1,y_2)Y'(y_1',y_2')Z(z_1,z_2,z_3)Z'(z_1',z_2',z_3')}
               =\sum_{x_3,x_3'}\tilde{B}_{xx_3y_1y_1'z_1z_1'}A_{x_3x_3'z_3z_3'}\tilde{B}_{x_3'x'y_2y_2'z_2z_2'}.
            \end{eqnarray}
            At the center, two $C$ tensors and one $\tilde{B}$ tensor form a new $C'$ tensor,
            \begin{eqnarray}
                \label{eq:newc1}
                \nonumber
                {C'}_{xx'Y(y_1,y_2,y_3)Y'(y_1',y_2',y_3')Z(z_1,z_2,z_3)Z'(z_1',z_2',z_3')}
                =\sum_{x_2,x_2'}C_{xx_2y_1y_1'z_1z_1'}\tilde{B}_{x_2x_2'y_2y_2'z_2z_2'}C_{x_2'x'y_3y_3'z_3z_3'}.
            \end{eqnarray}

\section{Work in progress and conclusions}
\label{sec:trgsign}
\begin{figure}[h]
 \begin{center}
 \includegraphics[width=3in]{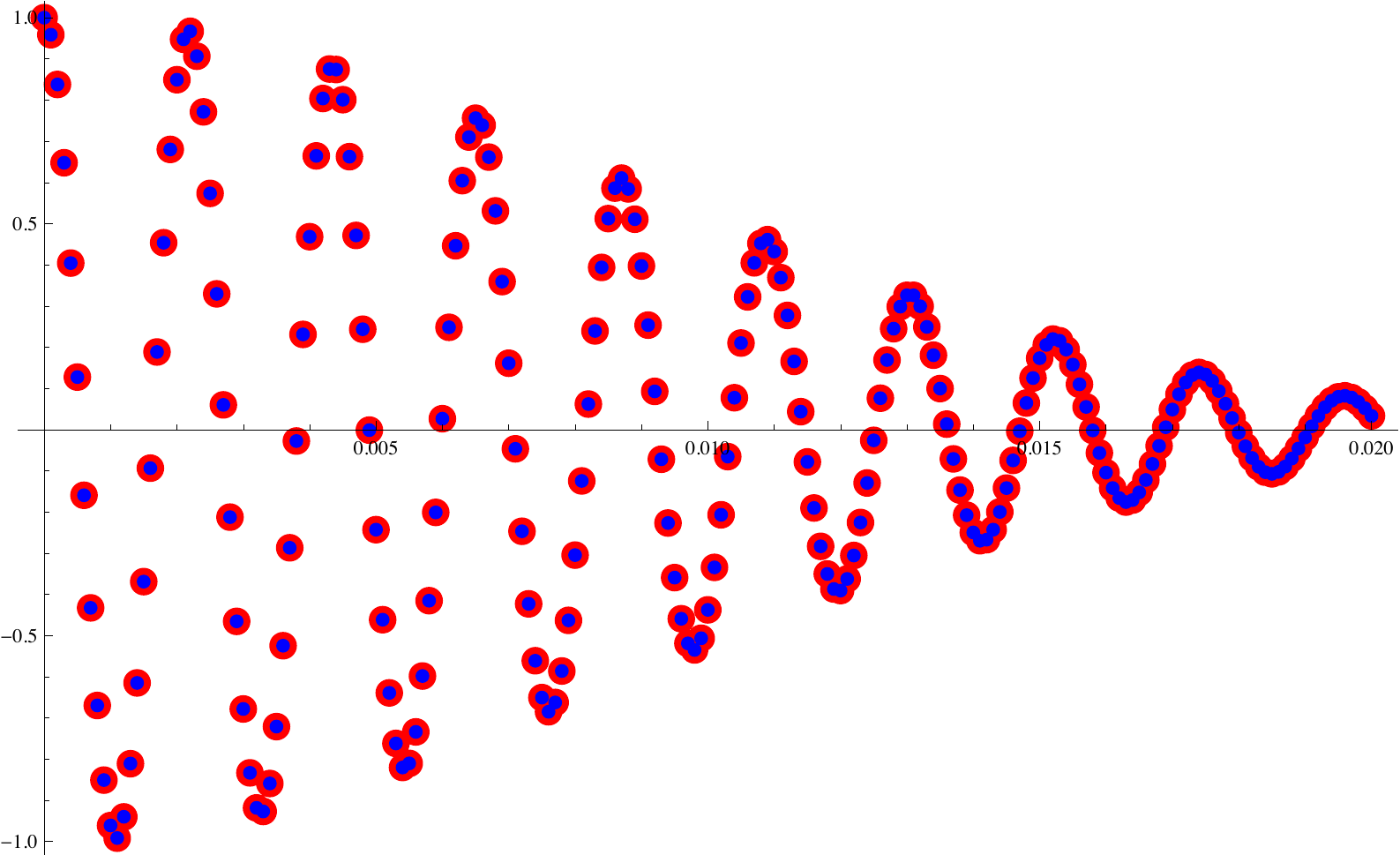}
 \caption{\label{fig:d20pp2} The real part of the partition function for $\beta =0.3+ix$ vs. x; red: result from HOTRG with 20 states; blue: exact solution (Onsager-Kaufmann). }
 \end{center}
 \end{figure}

\begin{figure}[h]
 \begin{center}
 \includegraphics[width=2.7in]{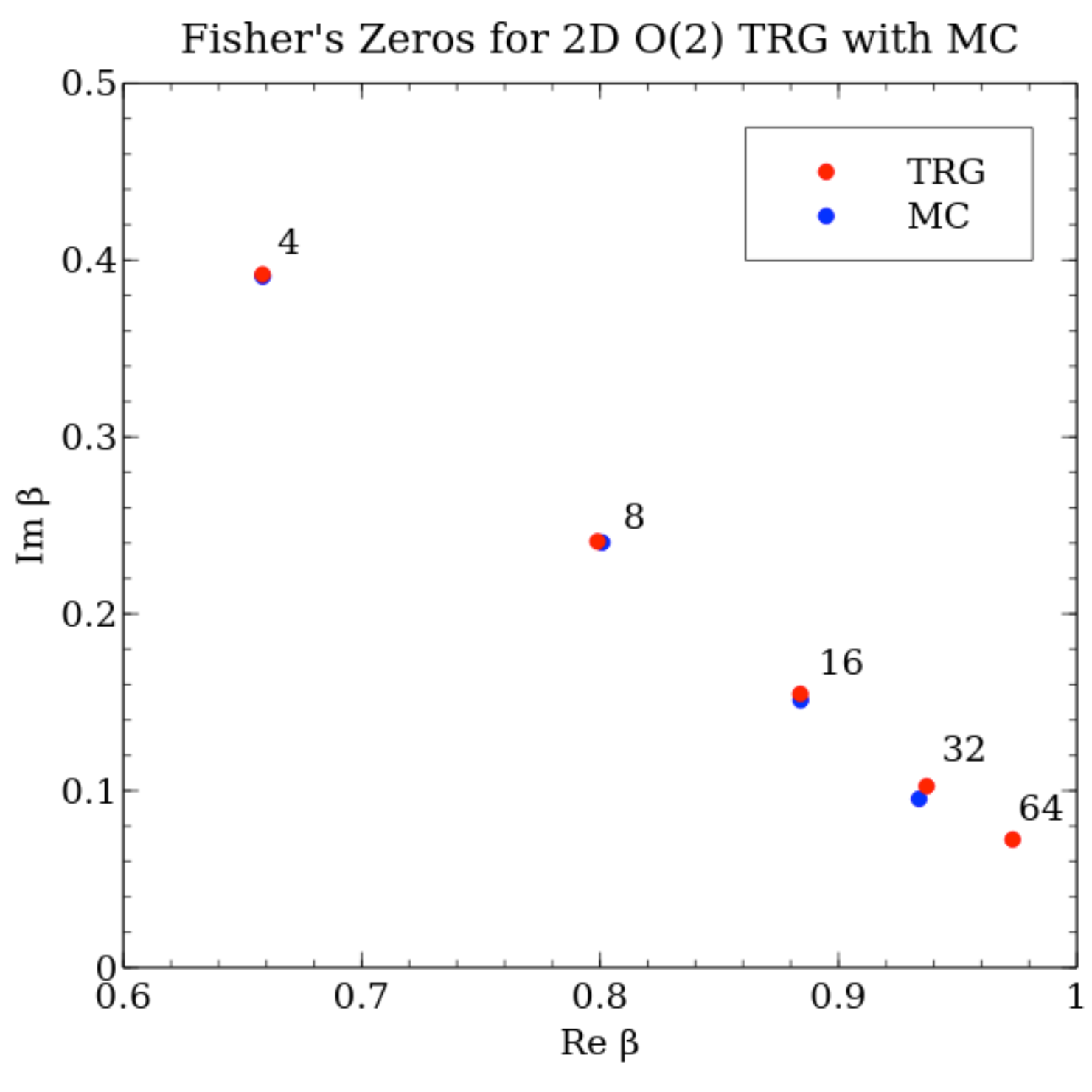}
 \caption{\label{fig:xyzeros} Fisher's zeros of XY model with $L=$ 4, 8, 16, 32, 64 for 30 states compared to MC}
 \end{center}
 \end{figure}

Successful numerical applications were mentioned at the conference and are now reported in recent preprints \cite{trgo2,trgsign}.
For the 2D Ising at complex $\beta$ (sign problem) and finite volume, the partition function obtained with the TRG has been compared to the exact Onsager-Kaufman solution. This is illustrated in Fig. \ref{fig:d20pp2}. In general, the TRG can reach much larger values of Im${\beta}$ than MC. The only regime where caution needs to be used is in the vicinity of Fisher's zeros where all the contributions tend to cancel and the results are more sensitive to truncation effects. This is explained at length in Ref. \cite{trgsign}. 
For the 2D $O(2)$ at finite volume, the Fisher's zero have been obtained and compared with those obtained by MC and reweighting. The good agreement is illustrated in Fig. \ref{fig:xyzeros}. 
The critical properties of the 2D $O(2)$ at large volume have been compared with MC \cite{trgo2}.
More recently, calculations for the 2D $O(2)$ with chemical potential (another sign problem) have been compared to results obtained with the worm algorithm  \cite{Banerjee:2010kc} and 
an excellent agreement was found \cite{shy}. We plan to study modulated phases in clock models with imaginary chemical potential and compare with results presented at the conference \cite{Meisinger:2013zfa}.

      In summary, the TRG method allows us to achieve the initial Wilsonian program (block spinning) in an exact way. 
      It applies to most classical lattice models. The HOTRG procedure allows controllable truncations. Successful numerical implementations at large volumes and for models with  sign problem are very promising. In general the implementation of the HOTRG involves computational demands that are very different from those of 
      conventional MC methods. Namely, memory considerations overwhelm CPU performance. On the analytical side, the somehow irregular behavior at intermediate 
      truncations remains \cite{efrati13} to be understood. Many new applications are possible and we expect that the TRG will become a useful complement of the 
      conventional MC methods.  
\vskip5pt
\noindent
{\bf Acknowledgements}

 This research was supported in part  by the Department of Energy
under Award Numbers DE-SC0010114 and FG02-91ER40664. 
We used resources of the National Energy Research Scientific Computing Center, which is supported by the Office of Science of the U.S. Department of Energy No. DE-AC02-05CH11231. 
Y. L. is supported by the URA Visiting Scholars' program. 
Fermilab is operated by Fermi Research Alliance, LLC, under Contract
No.~DE-AC02-07CH11359 with the United States Department of Energy.
 Our work on the subject started  while attending the KITPC workshop ``Critical Properties of Lattice Models" in summer 2012. 
 Y. M. did part of the work while at the workshop ``LGT in the LHC Era" in summer 2013 at the Aspen Center for Physics supported by NSF 1066293.

\end{document}